\documentclass[12pt]{article}
\usepackage{epsf}
\def\be{\begin{equation}}
\def\ee{\end{equation}}
\def\bea{\begin{eqnarray}}
\def\eea{\end{eqnarray}}

\begin{document}
\begin{titlepage}
\begin{center}
{\Large \bf William I. Fine Theoretical Physics Institute \\
University of Minnesota \\}
\end{center}
\vspace{0.2in}
\begin{flushright}
FTPI-MINN-19/06 \\
UMN-TH-3815/19 \\
February 2019 \\
\end{flushright}
\vspace{0.3in}
\begin{center}
{\Large \bf Radiative and pionic transitions $Z_c(4020)^0 \to X(3872) \gamma$ and  $Z_c(4020)^\pm \to X(3872) \pi^\pm$
\\}
\vspace{0.2in}
{\bf  M.B. Voloshin  \\ }
William I. Fine Theoretical Physics Institute, University of
Minnesota,\\ Minneapolis, MN 55455, USA \\
School of Physics and Astronomy, University of Minnesota, Minneapolis, MN 55455, USA \\ and \\
Institute of Theoretical and Experimental Physics, Moscow, 117218, Russia
\\[0.2in]

\end{center}

\vspace{0.2in}

\begin{abstract}
The radiative transitions from the neutral exotic $Z_c(4020)^0$ resonance to $X(3872)$ with emission of a photon and the pionic transitions from the charged $Z_c(4020)^\pm$ to $X(3872)$ and a charged pion $\pi^\pm$ are considered for both charmoniumlike states being charmed meson-antimeson molecules. The underlying processes for these transitions are the decays of the charmed vector meson to pseudoscalar plus a photon or a pion.  It is found that  the discussed transition rates typically amount to a branching fraction of several permil with a peak at  the vector meson pair threshold.
  \end{abstract}
\end{titlepage}

The charmoniumlike exotic resonance $X(3872)$ observed~\cite{bellex,cdfx,d0x,babarx} within a fraction of MeV from the threshold of the charmed meson pairs $D^{*0} D^0$ commands a considerable interest due to its rather exotic properties that very strongly suggest that this state is dominantly a four-quark object with a relatively small admixture of a $c \bar c$ charmonium. Furthermore, the four-quark component is likely dominated by a molecular $J^{PC}=1^{++}$ state made from $S$-wave $D^{*0} \bar D^0$ pairs and their charge conjugates. (A discussion of the arguments for such description can be found e.g. in the review~\cite{mvch}.) Most recenly this picture of  mostly exotic structure  of the $X(3872)$ state has received an additional boost from the observation\cite{besxchi} of the decay $X(3872) \to \chi_{c1} \pi^0$, which transitions would be very small for a pure charmonium~\cite{dv07} and are expected at about the observed level for a four-quark system~\cite{dv07,fm,mehen}, in particular for a $D^{*0} \bar D^0+ \bar D^{*0}  D^0$ molecule.

The subject of this paper are radiative and pionic transitions involving the $X(3872)$ state, namely the transitions to this resonance from the  $Z_c(4020)$ states. The $Z_c(4020)$ isotopic triplet of resonances with the quantum numbers $I^G(J^P)=1^+(1^+)$ observed~\cite{beszcc,beszcn} near the threshold of charmed vector meson pairs $D^* \bar D^*$ likely presents another example of a molecular state~\footnote{In the latest Particle Data Group review tables~\cite{pdg} the discussed charmoniumlike states are renamed as $\chi_{c1}(3872)$ and $X(4020)$. Here these resonances are referred to by their original names assigned at the discovery~\cite{bellex,beszcc}: $X(3872)$ and $Z_c(4020)$.}. (Recent reviews of the experimental and theoretical developments regarding the exotic so-called XYZ states can be found in Refs.~\cite{dsz,Guo17,Ali17}.) One can certainly notice that a radiative transition is only possible for the neutral component of the isotriplet: $Z_c(4020)^0 \to X(3872) \gamma$, while the pion emission is only possible from the charged resonance: $Z_c(4020)^\pm \to X(3872) \pi^\pm$, since emission of the neutral pion is excluded by the $C$ parity, and the $G$ parity violating charged pion emission is allowed due to large isospin breaking in $X(3872)$. 

The underlying processes for the considered here transitions are the decays $D^{*0} \to D^0 \gamma$ and $D^{*+} \to D^0 \pi^+$ (and their charge conjugates). The considered initial and final molecular states are very close to the respective thresholds, so that the motion of the mesons is dominated by the free one at distances beyond the range of the strong interaction (small interaction radius approximation). Therefore, the transition amplitudes can be calculated using the knowledge of the underlying decays for free on-shell mesons. A similar approach has been employed previously for calculations of the decays $X(3872) \to D^0 \bar D^0 \gamma$ and $X(3872) \to D^0 \bar D^0 \pi^0$~\cite{mv03} and of the process $e^+e^- \to X(3872) \gamma$~\cite{dv06} at energy near the $D^* \bar D^*$ threshold and  at the $Y(4260)$ peak~\cite{ghmwz}, with the latter process observed experimentally~\cite{besgx}. For free vector mesons the widths of the radiative and the pionic decays are  several tens of keV. Thus one can expect the rates of the discussed transitions between the molecular states to be in the same ballpark thus amounting to several permil of the total width of $Z_c(4020)$.

The position of the $X(3872)$ peak relative to the $D^{*0} D^0$ threshold is in fact uncertain. According to the Tables~\cite{pdg} $M_X = 3871.69 \pm 0.17\,$MeV, while $M(D^{*0})+ M(D^0) = 3871.68 \pm 0.10\,$MeV, so that it is not known whether it is a resonance, a bound state, or a peak corresponding to a virtual state~\cite{hkkn,mv07} in which case the position of the peak exactly coincides with the threshold. In what follows we shall consider $X(3872)$ as a shallow bound state with a small binding energy $\varepsilon = M(D^{*0})+ M(D^0) - M_X$ not exceeding few tenths of MeV. It appears that only under this assumption the discussed here radiative and pionic transitions have a sizable rate and also under this assumption there is a reasonable agreement between the measured yield of $X(3872) \gamma$ in $e^+e^-$ annihilation at 4.26\,GeV and the calculation in Ref.~\cite{ghmwz}.

The Fock decomposition of the wave function of $X(3872)$ can be generally written as
\be
X = a_0 \, \psi_0 + \sum_i a_i \, \psi_i~,
\label{fd}
\ee
where $\psi_0$ is the S-wave state of neutral charmed mesons $(D^0 \bar D^{*0} + \bar D^0  D^{*0})/\sqrt{2}$, and $\psi_i$ stand for `other' hadronic states including the heavier pair of charged mesons, $(D^+ D^{*-}+ D^- D^{*+})/\sqrt{2}$, pure $c \bar c$ charmonium, etc. Due to the extreme proximity to the threshold for neutral mesons, they move freely beyond the range of strong interaction, and thus their momentum space wave function in the rest frame of $X$ reads as
\be
\phi_n(\vec p) = {\sqrt{8 \pi \kappa} \over p^2 + \kappa^2}~,
\label{ff}
\ee
where $p$ stands  for $|\vec p|$, and the effective momentum scale $\kappa$ is determined by the binding energy $w$ and the reduced mass $m_r \approx 966\,$MeV in the $D^0 \bar D^{*0}$ system as $\kappa = \sqrt{2 m_r w} \approx 14\,{\rm MeV} \, \sqrt{w/0.1\,{\rm MeV}}$. The wave function in Eq.(\ref{ff}) is normalized to one, so that if used for $\psi_n$ in Eq.(\ref{fd}), the statistical weight for the  $(D^0 \bar D^{*0} + \bar D^0  D^{*0})/\sqrt{2}$ state inside $X(3872)$ is given by $|a_0|^2$. 

The presence of the component with the charged mesons can then be also described using the understanding that at short distances (large momenta) the strong interaction is isotopically neutral and the effects of the isospin violating mass difference between charged and neutral mesons should disappear~\cite{mv05,hkkn,mv07}. Approximating the charged meson pair wave function $\phi_c$ by that of the free motion and requiring that it coincides with $\phi_n$ at large $p$, one readily finds
\be
\phi_c(\vec p)= {\sqrt{8 \pi \kappa} \over p^2  + \kappa^2+ 2 m_r \Delta}~,
\label{fc}
\ee
where $\Delta = M(D^{*+}) + M(D^+)- M(D^{*0}) - M(D^0) \approx  8.2\,$MeV. Then the statistical weight of the charged mesons relative to that of the neutral ones is approximately given by $\sqrt{w / \Delta} \approx 0.11 \, \sqrt{w/0.1\,{\rm MeV}}$. In what follows we neglect any dynamic effects of the small component of the $X(3872)$ wave function with the pair of charged mesons as well as of other Fock states, and account for their presence only in the normalization factor $a_0$ for the neutral component.

The $Z_c(4020)$ resonances are considered here as dominantly coupled to $I^G(J^P) = 1^+(1^+)$ $S$-wave pairs of charmed vector meson and antimeson. In terms of the isotopic components this coupling can be written as
\be
L_{int} \left [ Z_c(4020)^0 \right ] = {\rm i} \, {h \over 2} \, \epsilon_{ljk} Z_l^0 \, \left (D_j^0 \bar D_k^0 - D_j^+ D_k^- \right )~,
\label{ln}
\ee
for the neutral $Z_c(4020)^0$ resonance, and
\be
L_{int} \left [ Z_c(4020)^\pm \right ] = {\rm i} \, {h \over \sqrt {2}} \, \epsilon_{ljk} Z_l^- \, \left ( D_j^+ \bar D_k^0 \right ) + {\rm h.c.}
\label{lc}
\ee
for the charged ones. The spatial vector indices $l,j,k$ label the components of the polarization amplitudes of the corresponding spin 1 resonances, and the non-relativistic normalization (to one) for the wave functions of heavy states is used throughout the present paper. The absolute value of the coupling constant $h$ determines the widths of the $Z_c(4020)$ resonances. Clearly, for these resonances the parameters of the neutral and the charged states can be somewhat different, e.g. due to the lower threshold for the neutral vector mesons than for the charged ones. Neither the mass splitting nor the difference of the widths for the $Z_c(4020)$ states is sufficiently studied experimentally. We use here for estimate the `average' values~\cite{pdg}, $M[Z_c(4020)] = 4024\,$ MeV and $\Gamma[Z_c(4020)] = 13\,$MeV, assuming that these parameters are for the charged state and that its total width is dominated by the decay into $D^{*+} \bar D^{*0}$. Then the constant $h$ can be estimated as
\be
|h|^2= {2 \pi \, \Gamma_Z  \over M(D^*) \, p_+} \approx { 1 \over 2.9\, {\rm GeV}}~,
\label{hv}
\ee
where $p_+ \approx 118\,$MeV is the vector meson momentum in the decay $Z_c(4020)^+ \to D^{*+} \bar D^{*0}$.

\begin{figure}[ht]
\begin{center}
 \leavevmode
    \epsfxsize=14cm
    \epsfbox{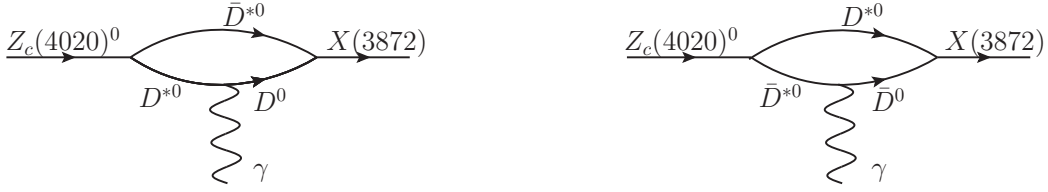}
    \caption{The graphs describing the radiative transition $Z_c(4020)^0 \to X(3872) \gamma$. }
\end{center}
\end{figure}

The amplitude of the transition $Z_c(4020)^0 \to X(3872) \gamma$ arises due to the processes $D^{*0} \to D^0 \gamma$ and  $\bar D^{*0} \to \bar D^0 \gamma$ as shown in Fig.~1. The strength of the photon coupling in these processes can be estimated from the data~\cite{pdg} on the relative rate of the radiative decay of $D^{*0}$ and the pionic decay $D^{*0} \to D^0 \pi^0$ and the isotopic relation of the latter decay to the charged meson decay $D^{*+} \to D^0 \pi^+$, whose absolute rate is known. In this way one readily estimates $\Gamma(D^{*0} \to D^0 \gamma) \approx 40\,$KeV. Then the transion magnetic moment $\mu$ in the $D^{*0} D^0 \gamma$ coupling ,
\be
L_\gamma = {\rm i} \, \mu \, \epsilon_{ij l} D_i^0 a_j k_l \, D^0 + {\rm h.c.}
\label{dmu}
\ee
with $\vec k$ being the photon momentum and $\vec a$ its polarization amplitude, can be estimated as 
\be
|\mu| = \sqrt {3 \pi \,  \Gamma(D^{*0} \to D^0 \gamma) \over \omega^3} \approx {1 \over 2.6 \,{\rm GeV}}~,
\label{vmu}
\ee
where $\omega = |\vec k|$ is the energy of the photon.\footnote{One can also notice that the contribution from a similar process with the charged charmed mesons, $D^{*\pm} \to D^\pm \gamma$ is suppressed by a significantly smaller radiative transition rate for the charged mesons than for the neutral~\cite{pdg} in addition to the discussed suppression of the statistical weight of the charged mesons in the wave function of $X(3872)$.}

The amplitude for the discussed radiative transition $Z_c(4020)^0 \to X(3872) \gamma$ at the total energy $E$ near the threshold $E_{thr}$ for the vector meson pair given by the graphs of Fig.~1 can be readily found using the nonrelativistic perturbation theory:
\be
A \left [ Z_c(4020)^0 \to X(3872) \gamma \right ] = \left [ (\vec Z \cdot \vec k) (\vec X \cdot \vec a) - 
(\vec Z \cdot \vec a) (\vec X \cdot \vec k) \right ] \, F_\gamma/ \sqrt{2}
\label{ag}
\ee
with  the factor $F_\gamma$ given as
\bea
&&F_\gamma = a_0 \, h \, \mu  \,  M(D^*) \, \int \,  {d^3p / (2 \pi)^3 \over \rho^2 + \vec p \, ^2 - {\rm i 0}} \, {\phi_n(\vec p - \vec k/2) + \phi_n (-\vec p - \vec k/2) \over 2 } = \nonumber \\
&& a_0 \, h \, \mu \,  M(D^*) \,  \sqrt{8 \pi \kappa} \, \int \,  {d^3p / (2 \pi)^3 \over (\rho^2 + \vec p\, ^2 - {\rm i} 0 ) [\kappa^2 + (\vec p - \vec k/2)^2] }~,
\label{fg}
\eea
where $\rho^2 = M(D^*) \, (E_{thr}-E)$ with $E_{thr}=2 M(D^{*0}) \approx 4013.7\,$MeV, and the infinitesimal imaginary shift ${\rm i} 0 $ defines the proper analytic continuation at negative $\rho^2$, corresponding to energy above the threshold. The integral in the latter expression is readily calculated with the result
\be
F_\gamma = { a_0 \, h \, \mu \,   M(D^*)  \sqrt{\kappa } \over \sqrt{2 \pi} \, \omega} \left [ \arctan {\omega^2 + 4 \rho^2 - 4 \kappa^2 \over 4 \omega \kappa} +  \arctan {\omega^2 + 4 \kappa^2 - 4 \rho^2 \over 4 \omega \rho} \right ] ~.
\label{fgr}
\ee 
The transition rate is given in terms of $F_\gamma$ as 
\be
\Gamma \left [ Z_c(4020)^0 \to X(3872) \gamma \right ] = {|F_\gamma|^2 \omega^3 \over 3 \pi}~.
\label{rg}
\ee

\begin{figure}[ht]
\begin{center}
 \leavevmode
    \epsfxsize=7.5cm
    \epsfbox{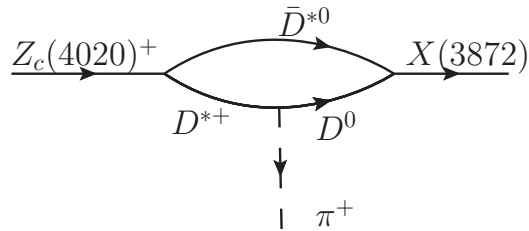}
    \caption{The mechanism for the pionic transition $Z_c(4020)^+ \to X(3872) \pi^+$. }
\end{center}
\end{figure}

The treatment of the  pionic transition $Z_c(4020)^+ \to X(3872) \pi^+$ follows the same lines as of the radiative one. The mechanism for this transition is shown in Fig.~2, and the main difference from the photon case is the $D^* D \pi$ coupling described by the interaction
\be
L_{\pi} = {g \sqrt{2} \over f_\pi} \, (\vec  D^+ \cdot \vec q) D^0 \pi^-~,
\label{lpi}
\ee
where $\vec q$ is the pion momentum, the pion decay constant $f_\pi \approx 132\,$MeV is used for normalization. In this notation the value of the dimensionless coupling constant found from the experimental width~\cite{pdg} of the $D^{*+}$ decay is $g^2 \approx 0.15$. Using the expression (\ref{lpi}) and the formula in Eq.(\ref{lc}) we find that the amplitude arising from the mechanism of Fig.~2 can be written as
\be
A \left [ Z_c(4020)^+ \to X(3872) \pi^+ \right ] = \epsilon_{ijk} Z_i X_j q_k \, F_\pi/ \sqrt{2}
\label{api}
\ee
with the factor $F_\pi$ determines the transition rate:
\be
\Gamma \left [ Z_c(4020)^+ \to X(3872) \pi^+ \right ] = {|F_\pi|^2 q^3 \over 6 \pi}~.
\label{rpi}
\ee
and is given by
\bea
&&F_\pi = a_0 \, h \, {g \over f_\pi}  \,  M(D^*) \, \int \,  {d^3p / (2 \pi)^3 \over \rho^2 + \vec p \, ^2 - {\rm i 0}} \, \phi_n(\vec p - \vec q / 2)  = \nonumber \\
&& a_0 \, h \, {g \over f_\pi} \,  M(D^*) \,  \sqrt{8 \pi \kappa} \, \int \,  {d^3p / (2 \pi)^3 \over (\rho^2 + \vec p\, ^2 - {\rm i} 0 ) [\kappa^2 + (\vec p - \vec q/2)^2] } = \nonumber \\
&&{ a_0 \, h \, g \,   M(D^*)  \sqrt{\kappa } \over \sqrt{2 \pi} \, f_\pi \, q} \left [ \arctan {q^2 + 4 \rho^2 - 4 \kappa^2 \over 4 q \kappa} +  \arctan {q^2 + 4 \kappa^2 - 4 \rho^2 \over 4 q \rho} \right ]~.
\label{fpi}
\eea
Here $q$ stands for $|\vec q|$ and, as before, the notation $\rho^2 = M(D^*) \, (E_{thr}-E)$ is used, but in this case $E_{thr} = M(D^{*+})+ M(D^{*0}) \approx 4017.1\,$MeV is the threshold for the $D^{*+} + \bar D^{*0}$ vector meson pair.\footnote{It can be noted that the discussed here derivation of the transition amplitudes using the standard nonrelativistic perturbation theory and essentially following the lines of Ref.~\cite{dv06} is fully equivalent to a calculation~\cite{ghmwz} based on nonrelativistic limit of  relativistic Feynman graphs of Fig.~1 and using the Weinberg formula~\cite{Weinberg} for the coupling of a shallow bound state to unbound states in the continuum.} 

One can readily see that according to the formulas (\ref{fg}), (\ref{fgr}) for the radiative transition and (\ref{rpi}), (\ref{fpi}) for the emission of the pion, the discussed rates should exhibit a significant variation with energy $E$ near the threshold. It thus appears reasonable to introduce an energy dependent branching fraction, $r(E) = \Gamma(E)/\Gamma$, with $\Gamma$ being the total width of the resonance. Clearly, this ratio describes the relative significance of the discussed transitions both on and off the nominal position of the resonance. In addition to the uncertainty in the parameters of the $Z_c(4020)$ resonances, specific numerical estimates from the discussed formulas obviously suffer from the uncertainty in the parameter $\kappa$ determined by the binding energy $w$ in the $X(3872)$ state, and also from poor knowledge of the statistical weight $|a_0|^2$ of the molecular $(D^0 \bar D^{*0} + \bar D^0  D^{*0})/sqrt{2}$ component. Accordingly, we consider few `representative' values for $w$: 0.1, 0.2 and 0.4\,MeV, and plot the `normalized' energy dependent branching fraction $R(E) = r(E)/|a_0|^2$ for the radiative transition, $R_\gamma(E)$ in Fig.~3, and for the pionic process, $R_\pi(E)$ in Fig.~4. 

\begin{figure}[ht]
\begin{center}
 \leavevmode
    \epsfxsize=7.5cm
    \epsfbox{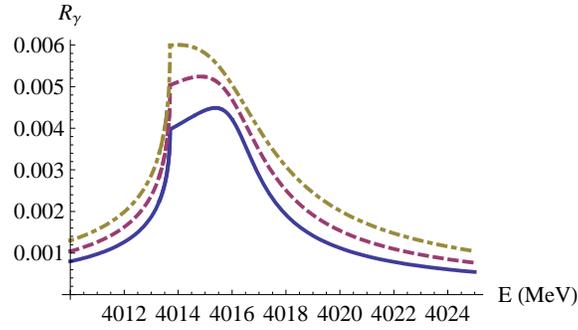}
    \caption{The normalized energy dependent branching fraction for the radiative transition $Z_c(4050)^0 \to X(3872) \gamma$ at the values of the binding energy $w$: 0.1\,MeV (solid), 0.2\,MeV (dashed) and 0.4\,MeV (dotdashed). }
\end{center}
\end{figure} 

\begin{figure}[ht]
\begin{center}
 \leavevmode
    \epsfxsize=7.5cm
    \epsfbox{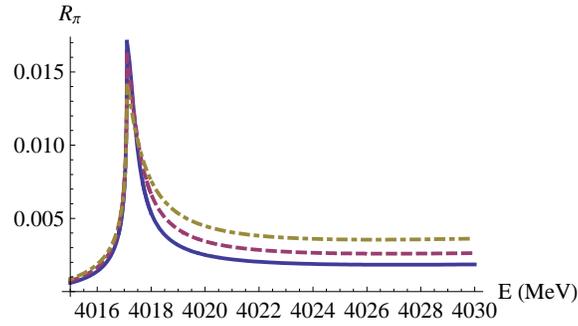}
    \caption{The normalized energy dependent branching fraction for the pionic transition $Z_c(4020)^\pm \to X(3872) \pi^\pm$ at the values of the binding energy $w$: 0.1\,MeV (solid), 0.2\,MeV (dashed) and 0.4\,MeV (dotdashed). }
\end{center}
\end{figure}

To summarize. In the scenario where  $X(3872)$ is dominantly an $S$-wave $(D^0 \bar D^{*0} + \bar D^0  D^{*0})/\sqrt{2}$ molecule and the $Z_c(4020)$ states is an isotopic triplet of near-threshold $S$-wave $D^* \bar D^*$ resonances, there should be transitions $Z_(4020)^0 \to X(3872) \gamma$ and $Z_c(4020)^\pm \to X(3872) \pi^\pm$ induced by the free-meson processes $D^{*0} \to D^0 \gamma$ and $D^{*+} \to D^0 \pi^+$. The amplitudes of these transitions are calculable in terms of the molecular content of $X(3872)$ and are given by Eqs.~(\ref{fg}) and (\ref{fpi}).  The calculated rates are quite challenging for experimental observation and correspond to at most several tenths percent in terms of the branching fraction for the $Z_c(4020)$ resonances.  However an observation of one or both of the discussed processes should allow an important quantitative study of the molecular picture.

This work is supported in part by U.S. Department of Energy Grant No.\ DE-SC0011842.

\end{document}